\documentclass[aps,prd,twocolumn,showpacs,nofootinbib]{revtex4-1}

\usepackage{amssymb} \usepackage{amsmath} \usepackage{graphicx}
\usepackage{epsfig,latexsym}

\RequirePackage{xspace} \allowdisplaybreaks

\begin{document}

\def\bef{\begin{figure}}
\def\eef{\end{figure}}

\newcommand{\nl}{\nonumber\\}

\newcommand{\ans}{ansatz }
\newcommand{\be}[1]{\begin{equation}\label{#1}}
\newcommand{\beq}{\begin{equation}}
\newcommand{\ee}{\end{equation}}
\newcommand{\beqn}[1]{\begin{eqnarray}\label{#1}}
\newcommand{\eeqn}{\end{eqnarray}}
\newcommand{\bd}{\begin{displaymath}}
\newcommand{\ed}{\end{displaymath}}
\newcommand{\mat}[4]{\left(\begin{array}{cc}{#1}&{#2}\\{#3}&{#4}
\end{array}\right)}
\newcommand{\matr}[9]{\left(\begin{array}{ccc}{#1}&{#2}&{#3}\\
{#4}&{#5}&{#6}\\{#7}&{#8}&{#9}\end{array}\right)}
\newcommand{\matrr}[6]{\left(\begin{array}{cc}{#1}&{#2}\\
{#3}&{#4}\\{#5}&{#6}\end{array}\right)}
\newcommand{\cvb}[3]{#1^{#2}_{#3}}
\def\lsim{\raise0.3ex\hbox{$\;<$\kern-0.75em\raise-1.1ex
e\hbox{$\sim\;$}}}
\def\gsim{\raise0.3ex\hbox{$\;>$\kern-0.75em\raise-1.1ex
\hbox{$\sim\;$}}}
\def\abs#1{\left| #1\right|}
\def\simlt{\mathrel{\lower2.5pt\vbox{\lineskip=0pt\baselineskip=0pt
           \hbox{$<$}\hbox{$\sim$}}}}
\def\simgt{\mathrel{\lower2.5pt\vbox{\lineskip=0pt\baselineskip=0pt
           \hbox{$>$}\hbox{$\sim$}}}}
\def\unity{{\hbox{1\kern-.8mm l}}}
\newcommand{\eps}{\varepsilon}
\def\ep{\epsilon}
\def\ga{\gamma}
\def\Ga{\Gamma}
\def\om{\omega}
\def\omp{{\omega^\prime}}
\def\Om{\Omega}
\def\la{\lambda}
\def\La{\Lambda}
\def\al{\alpha}
\newcommand{\ov}{\overline}
\renewcommand{\to}{\rightarrow}
\renewcommand{\vec}[1]{\mathbf{#1}}
\newcommand{\vect}[1]{\mbox{\boldmath$#1$}}
\def\tm{{\widetilde{m}}}
\def\mcirc{{\stackrel{o}{m}}}
\newcommand{\Dm}{\Delta m}
\newcommand{\dm}{\varepsilon}
\newcommand{\tanb}{\tan\beta}
\newcommand{\nbar}{\tilde{n}}
\newcommand\PM[1]{\begin{pmatrix}#1\end{pmatrix}}
\newcommand{\up}{\uparrow}
\newcommand{\down}{\downarrow}
\def\omE{\omega_{\rm Ter}}
%
%%%%%%%%%%     mauri    %%%%%%%%%%%%%%%%%%%%%%%%%%%%%%%%%

\newcommand{\Dsusy}{{susy \hspace{-9.4pt} \slash}\;}
\newcommand{\DCP}{{CP \hspace{-7.4pt} \slash}\;}
\newcommand{\mc}{\mathcal}
\newcommand{\gr}{\mathbf}
\renewcommand{\to}{\rightarrow}
\newcommand{\gtc}{\mathfrak}
\newcommand{\wh}{\widehat}
\newcommand{\br}{\langle}
\newcommand{\kt}{\rangle}

%%%%%%%%%%%%%%%%%%%%%%%%%%%%%%%%%%%%%%%%%%%%%%%%%%%%%%%%%%

% barbara Ricci  %definizione di minore e maggiore simile
\def\lsim{\mathrel{\mathop  {\hbox{\lower0.5ex\hbox{$\sim$}
\kern-0.8em\lower-0.7ex\hbox{$<$}}}}}
\def\gsim{\mathrel{\mathop  {\hbox{\lower0.5ex\hbox{$\sim$}
\kern-0.8em\lower-0.7ex\hbox{$>$}}}}}
%%%%%%%%%%%%%%%%%%%%%%%%%%%%%%%%%%

\def\nn{\\  \nonumber}
\def\de{\partial}
\def\brf{{\mathbf f}}
\def\bbf{\bar{\bf f}}
\def\bF{{\bf F}}
\def\bbF{\bar{\bf F}}
\def\bA{{\mathbf A}}
\def\bB{{\mathbf B}}
\def\bG{{\mathbf G}}
\def\bI{{\mathbf I}}
\def\bM{{\mathbf M}}
\def\bY{{\mathbf Y}}
\def\bX{{\mathbf X}}
\def\bS{{\mathbf S}}
\def\bb{{\mathbf b}}
\def\bh{{\mathbf h}}
\def\bg{{\mathbf g}}
\def\bla{{\mathbf \la}}
\def\bmu{\mathbf m }
\def\by{{\mathbf y}}
\def\bmu{\mbox{\boldmath $\mu$} }
\def\bsig{\mbox{\boldmath $\sigma$} }
\def\bunity{{\mathbf 1}}
\def\cA{{\cal A}}
\def\cB{{\cal B}}
\def\cC{{\cal C}}
\def\cD{{\cal D}}
\def\cF{{\cal F}}
\def\cG{{\cal G}}
\def\cH{{\cal H}}
\def\cI{{\cal I}}
\def\cL{{\cal L}}
\def\cN{{\cal N}}
\def\cM{{\cal M}}
\def\cO{{\cal O}}
\def\cR{{\cal R}}
\def\cS{{\cal S}}
\def\cT{{\cal T}}
\def\eV{{\rm eV}}
%
%%%%%%%%%%%%%%%%%%%%%%%%%%%%%%%%%%%%%

\title{Direct generation of a Majorana mass for the Neutron from Exotic Instantons}

\author{Andrea Addazi$^1$}\email{andrea.addazi@infn.lngs.it}
\affiliation{$^1$ Dipartimento di Fisica,
 Universit\`a di L'Aquila, 67010 Coppito AQ and
LNGS, Laboratori Nazionali del Gran Sasso, 67010 Assergi AQ, Italy}

\begin{abstract}

We discuss a new mechanism 
in which non-perturbative quantum gravity effects 
directly generate
 a Majorana mass 
for the neutron. In particular, in string theory, 
exotic instantons can generate an effective six quark operator
by calculable mixed disk amplitudes. 
In a low string scale scenario,
with $M_{S}\simeq 10\div 10^{5}\, \rm TeV$, 
a neutron-antineutron oscillation can be reached 
in the next generation of experiments. 
we argue that protons and neutralinos are not destabilized and that dangerous FCNCs are not generated. 
We show an example of quiver theories, locally free by tadpoles 
and anomalies, 
reproducing MSSM plus a Majorana neutron and 
a Majorana neutrino.
These models naturally provide a viable baryogenesis mechanism 
by resonant RH neutrino decays, as well as a stable WIMP-like dark matter. 
%We also discuss how LHC data will provide useful
%inputs for our model, in the immediate future, 
%testing several different regions of the parameter space. 

%Such a mechanism represents a simple and calculable counter-example to the Wilsonian 
%UV completion of effective non-renormalizable operators.

\end{abstract}

\maketitle

\section{Introduction}

Testing low energy B/L-violating processes
is crucially important for our understanding 
of particle masses and matter-antimatter asymmetry 
in our Universe. 
Inspired by these deep motivations,
experiments on neutrinoless-double-beta-decays 
are very active, constraining the Majorana neutrino.
However, 
also a {\it neutron} can also have an effective Majorana mass term!
Majorana himself first suggested 
the neutron as a Majorana fermion \cite{Majorana37}.
In terms of Weinberg effective operators, 
such a mass term corresponds to 
a six-quark dimension-9 term
$(udd)^{2}/\mathcal{M}_{n\bar{n}}^{5}$.
$\mathcal{M}_{n\bar{n}},\tau_{n\bar{n}},\delta m_{n\bar{n}}$ 
are connected
 each other 
\be{connession}
\delta m_{n\bar{n}}=\tau_{n\bar{n}}^{-1}\simeq \left(\frac{\Lambda_{QCD}^{6}}{\mathcal{M}_{n\bar{n}}^{5}} \right)\simeq 10^{-25}\left(\frac{1000\, \rm TeV}{\mathcal{M}_{n\bar{n}}}\right)^{5}eV
\ee
where $\delta m_{n\bar{n}}$ is the Majorana mass of the neutron, $\tau_{n\bar{n}}$ the neutron-antineutron transition time \footnote{The validity of these estimations of non-perturbative QCD effects were checked 
in \cite{Classification,128,134,Buchoff:2015qwa}. } \footnote{Neutron-Antineutron transition can be 
also a hint for a new fifth force interaction  \cite{Addazi:2015pia}.}
Contrary to neutrini, a Majorana neutron can be directly tested in oscillations: neutron-antineutron 
transitions!
The best limit on neutron-antineutron transitions 
in vacuum is around $\tau_{n\bar{n}}>3\, \rm yr$,
from Baldo-Coelin experiment in Grenoble ('97') \cite{Baldo}.
This seems surprising if compared to 
other rare processes like proton decay 
$\tau_{p}>10^{35}\, \rm yr$ and 
neutrinoless double-beta decays $\tau_{0\nu\beta\beta}>10^{25}\,\rm yr$
\footnote{A so fast $n-\bar{n}$ transition in vacuum not destabilizes nuclei:
Neutron-Antineutron oscillations  
 are not just excluded by Superkamiokande experiment \cite{Abe:2011ky}.
In fact, contrary to decay processes, a $n-\bar{n}$ oscillation 
is strongly suppressed by the nuclear binding energy in the nuclear environment. 
In fact, a transition from a binding neutron 
to a practically unbounded antineutron inside nuclei is energetically unfavored: the effective low energy Hamiltonian 
of the neutron-antineutron system will have 
diagonal terms with a difference between of $10^{33}$ digits up than off-diagonal Majorana masses,
so that the transition time in nuclei will be suppressed up to $\tau_{n\bar{n}}^{Nuclei}>\tau_{exp}^{Nuclei}\simeq 10^{32}\, \rm yr$ \cite{Reviews1,Reviews2,Reviews3,Abe:2011ky}. See Appendix C for explicit calculations of these suppression effects. }.
As discussed in \cite{Reviews1,Reviews2,Reviews3}, there is the exciting opportunity to 
enhance current limits on $n-\bar{n}$ transitions up to $\tau_{n\bar{n}}\simeq 300\, \rm yr$,
with an experimental set-up {\it a la} Baldo-Coelin -with external magnetic field $|\vect{B}|\sim 10^{-5}\div 10^{-6}\, \rm Gauss$
\footnote{To realize these experiments with suppressed magnetic fields is necessary in order to 
not suppress $n-\bar{n}$ oscillations. For example, the Earth magnetic field ($0.5\, \rm Gauss$) 
will split the energies of neutron and antineutron of $2\mu_{n}B_{E}\simeq 10^{-11}\, \rm eV$.
This energy is $10^{12}$ higher than the present limit on the Majorana mass for the neutron 
($\delta m_{n\bar{n}}<10^{-23}\, \rm eV\simeq 10^{-8}s^{-1})$. }.

Recently, we have suggested that a Majorana mass for the neutron can be indirectly generated 
by non-perturbative effects of string theories known as exotic instantons
  \cite{Addazi:2014ila,Addazi:2015ata,Addazi:2015rwa,Addazi:2015hka,Addazi:2015eca,Addazi:2015fua,Addazi:2015oba}.
Exotic instantons are peculiarly different from gauge instantons. 
In fact, they cannot be reconstructed by an ADHM classification
of gauge instantons. Usually
gauge instantons can 'strongly' violate axial global symmetries.
On the other hand, exotic instantons can 'strongly' violate  global
vector-like symmetries, like Baryon/Lepton ones. These effects are often calculable and controllable
in open string theories. 
 In open-string theories, all instantons admit a simple "geometric" interpretation.
 In fact they are Eucliden D-branes, or E-branes, wrapping an internal cycle, that could intersect the `physical' D-branes.
 All gauge instantons of ADHM can be re-obtained by E-branes wrapping the same n-cycles of ordinary 
 D-branes. On the other hand, 'exotic instantons' are E-branes wapping {\it different} n-cycles of ordinary 
 D-branes. See 
 \cite{Bianchi:2009ij,Bianchi:2012ud,BIMP} for useful references on these aspects . 
As shown in  \cite{Addazi:2014ila,Addazi:2015ata,Addazi:2015rwa,Addazi:2015hka,Addazi:2015eca,Addazi:2015fua,Addazi:2015oba}, they can dynamically violate starting discrete symmetries $Z_{k}$ of a perturbative lagrangian. 
However, depending on their intersections with ordinary D-branes, 
they induce only specific operators, not necessarily  
all the possible $Z_{k}$-violating ones. 
In our first models, we have suggested that a $\mathcal{O}_{n\bar{n}}$ can be mediated 
by exotic matter with non-perturbative couplings induced by exotic instantons. 
In this sense, we have defined these mechanisms 'indirect' ones. 

In this paper, we suggest a new mechanism for the generation 
of an effective Majorana mass for the neutron:
we propose a simple and calculable mechanism generating $\mathcal{O}_{n\bar{n}}$
{\it totally} by exotic instantons, without the need for any mediator-fields.  
This could be a counter-example to UV Wilsonian completion 
of effective operators. Wilsonian UV completion has manifested itself as a successful 
approach in a lot of well understood examples in particle physics.
Probably the most famous example is the UV completion of the Fermi
model of weak interaction with a gauge theory of electroweak interactions, 
{\it i.e} the non-renormalizable four fermion interaction
is resolved as an exchange of a massive $W$ boson.
However, we will show how effective operators as six-quark operators can be completed by 
exotic instantons rather than by integrating-out new massive fields.
In our proposal, we will show how for $M_{S}=10\div 10^{5}\, \rm TeV$, 
a neutron-antineutron transition can be found in the next generation of experiments. 
We will see how this scenario can be compatible with two of the most elegant solutions to the hierarchy problem
of the Higgs mass:  TeV-scale susy,
or alternatively with low scale string theories with
 $M_{S}=10\, \rm TeV$ \cite{ADD1,AADD,RS1}. We are in LHC era, which will 
 provide a lot of important inputs also for our model in the immediate future!
 On the other hand, we will also consider 'less appealing' scenarios for LHC,
 compatible with a PeV-ish $\mathcal{M}_{n\bar{n}}$ . In fact, the case of a 
 $M_{SUSY}\simeq M_{S}\simeq 10^{3}\, \rm TeV$ remains compatible 
 with a $\mathcal{M}_{n\bar{n}}\simeq 1000\, \rm TeV$ neutron-antineutron transition.
 In this last case, electroweak scale is fine-tuned for a factor $m_{H}/M_{S}\sim 10^{-4}$,
 rather than $m_{H}/M_{Pl}\sim 10^{-17}$, {\it i.e} the hierarchy problem is alleviated. 
 We will also discuss explicit examples of un-oriented quiver theories
 reproducing at low energy limit a (MS)SM with a Majorana neutron, 
 a electroweak-scale mu-term, neutrini masses from Right-handed neutrini (compatible with RH-neutrino decayRal baryogenesis), WIMP-like dark matter, 
 without destabilizing nucleons. 

\section{(MS)SM quivers for an Exotic Majorana neutron}

\begin{figure}[t]
\centerline{ \includegraphics [height=6cm,width=0.7 \columnwidth]{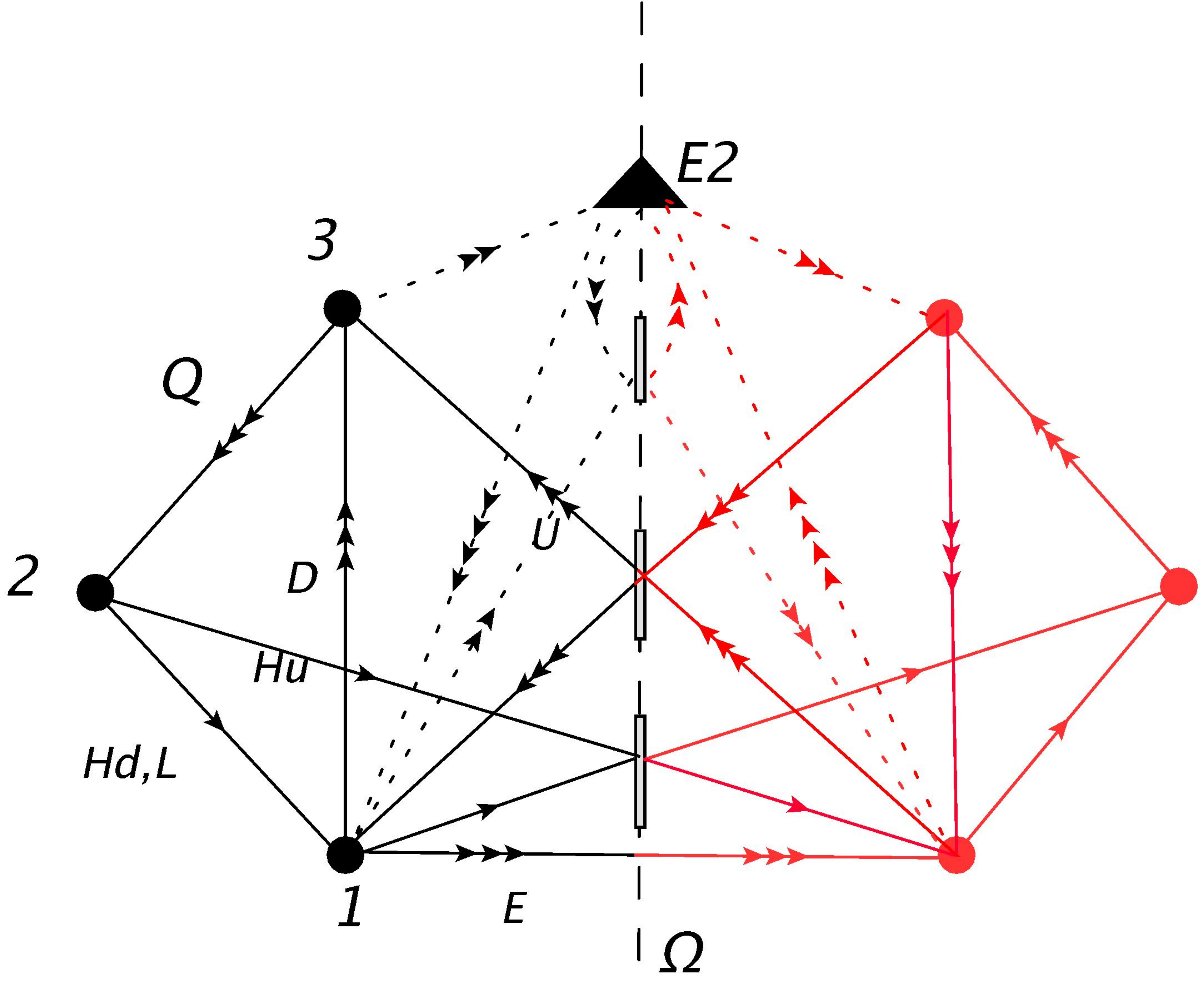}}
\vspace*{-1ex}
\caption{A tree-nodes quiver for a Majorana neutron. We not reports all instantons generating 
Yukawa couplings $y_{l}H_{d}LE^{c}$ and $y_{d}H_{d}QD^{c}$, and the $\mu$-term.
We also omit possible exotic fields getting consistent our quiver from the point
of view of tadpoles cancellations and massless hypercharge $U(1)_{Y}$, discussed in \cite{DB18}.
This quiver shows the number of intersections of $E2$ with 3,2,1-stacks,
generating six-quarks superpotentials. 
}
\label{plot}   % \ref{plot}
\end{figure}
\begin{figure}[t]
\centerline{ \includegraphics [height=6cm,width=0.7 \columnwidth]{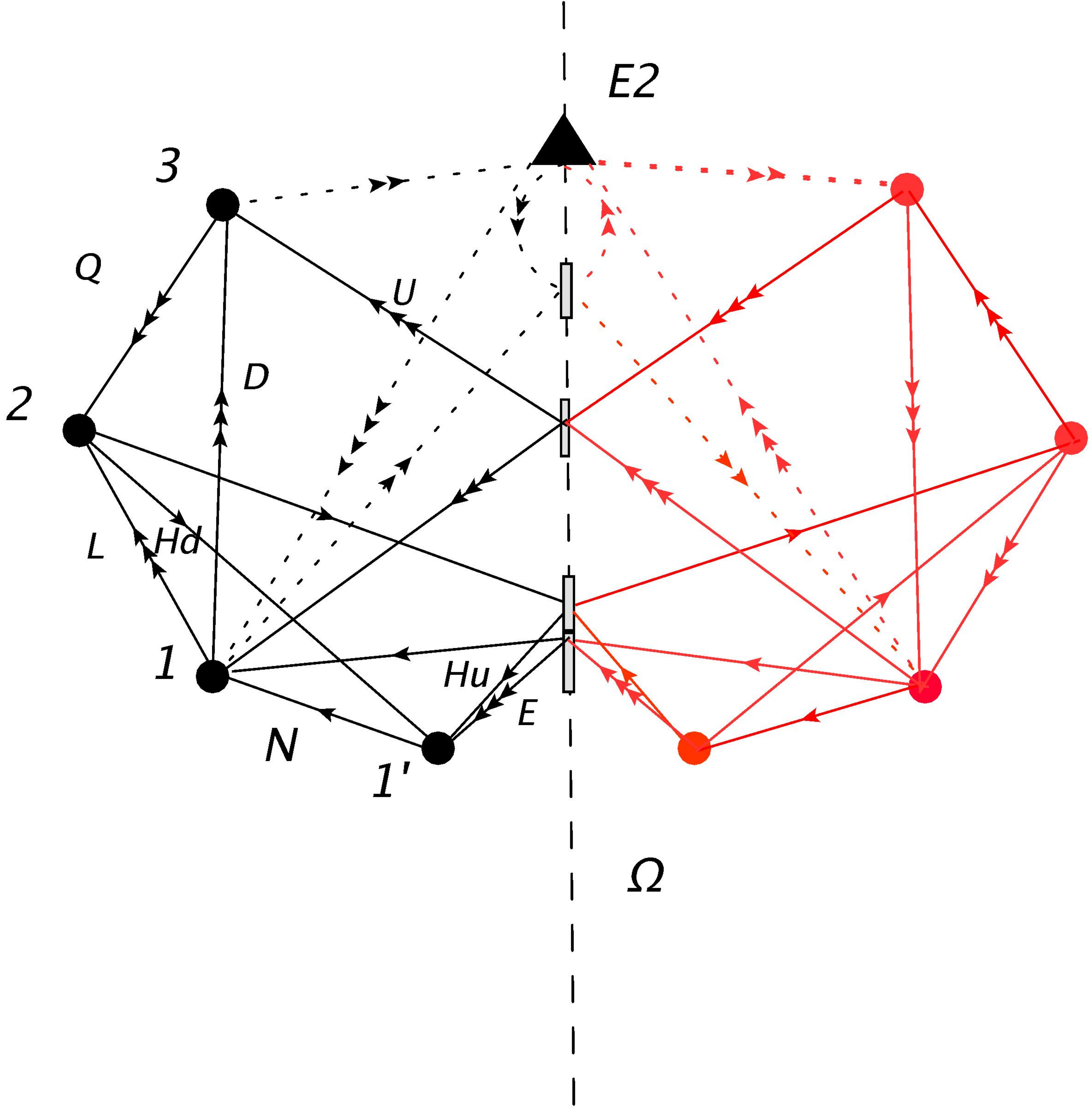}}
\vspace*{-1ex}
\caption{A four-nodes Madrid quiver for a Majorana neutron. We not reports all instantons generating 
Yukawa couplings, $\mu$-term and RH neutrini masses.
We also omit possible exotic fields getting consistent our quiver from the point
of view of tadpoles cancellations and massless hypercharge $U(1)_{Y}$, discussed in \cite{DB18}.
This quiver shows the number of intersections of $E2$ with 3,2,1-stacks,
generating six-quarks superpotentials. }
\label{plot}   % \ref{plot}
\end{figure}
\begin{figure}[t]
\centerline{ \includegraphics [height=6cm,width=0.7 \columnwidth]{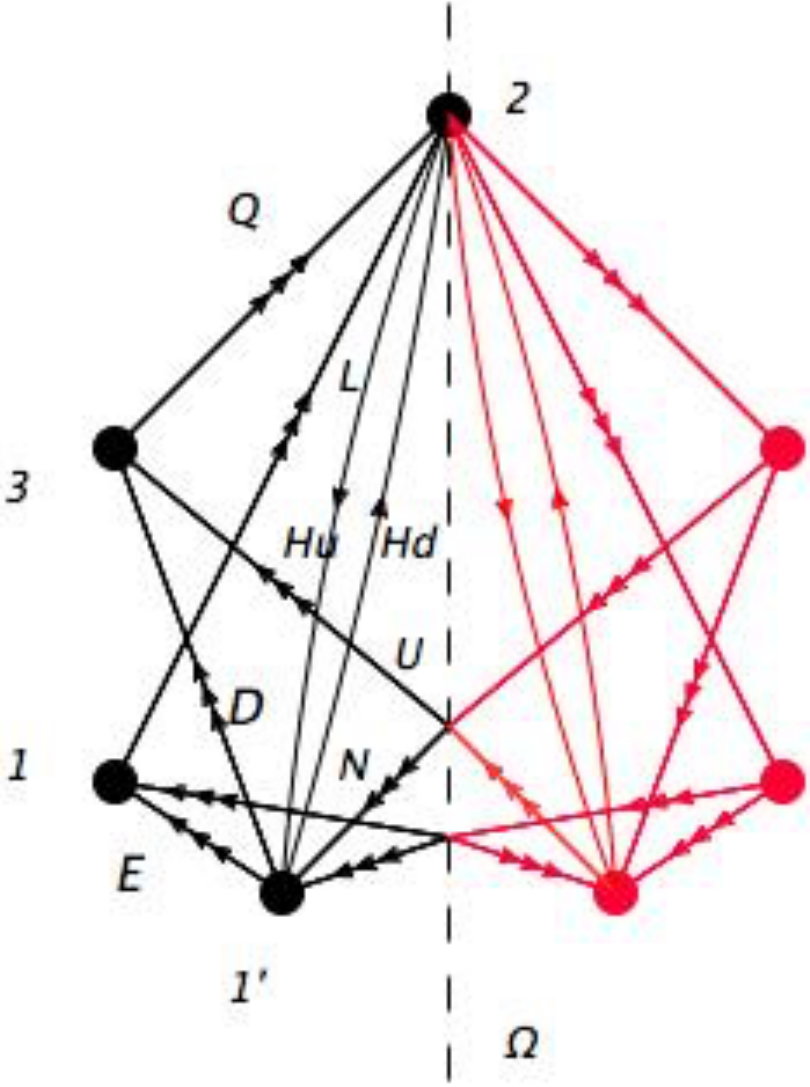}}
\vspace*{-1ex}
\caption{A four-nodes SM quiver. We omit the E2-instanton generating a Majorana mass for the neutron.
Similarly to Fig.1-2 such an $E2$-instanton can be consistently introduced. 
}
\label{plot}   % \ref{plot}
\end{figure}

A MSSM can be embedded in a quiver theory \footnote{See for useful references on open string theories and orientifolds
%Open Strings
\cite{Sagnotti1,Sagnotti2,Sagnotti3,Sagnotti6,Sagnotti7,Sagnotti8,Bianchi:1990yu,Bianchi:1990tb,Bianchi:1991eu,Angelantonj:1996uy,Angelantonj:1996mw}.
See \cite{AIQU,CSU,BKLS,DB1,DB2,DB3,DB4,DB5,DB8,DB18,DB10,DB12,DB14,DB16,DB17,DB22,DB23}
for useful literature on MSSM quiver theories.}. 
 with at least three nodes, 
reproducing a gauge theory $U(3)_{a}\times U(2)_{b}\times U(1)_{c}$,
in type II A string theory. 
The basic elements are few ones: $D6$-branes wrapping 3-cycles in the Calabi-Yau 
$CY_{3}$, one of these will be a flavor brane, one $\Omega$-plane, $E2$-instantons, open strings attached to 
$D6$-branes and $E2$-branes. 
Stacks of three parallel D6-branes will reproduce $U(3)_{a}$, 
and so on. (MS)SM matter fields in the bi-fundamental representations 
of SM gauge groups
are reproduced, in the low energy limit, 
by open (un)oriented strings attached to 
two intersecting D6-branes' stacks. 
For example, a $Q_{L}$ superfield 
comes from an open string attached to 
a-stack, reproducing $U(3)$, and 
b-stack, reproducing $U(2)$. 
The number of generations is reproduced by 
the number of intersections among D6-branes' stacks.
For example, the three generations of quarks 
correspond to three intersections between a-stack and b-stack.
In these models, hypercharge $U(1)_{Y}$ is a massless combination  
of $U(1)_{a}$ contained in $U(3)_{a}\simeq SU(3)\times U(1)_{a}$;
$U(1)_{b}$ contained in $U(2)_{b}\simeq SU(2)\times U(1)_{b}$; 
and $U(1)_{c}$:
\be{hypercharge}
U(1)_{Y}=q_{a}U(1)_{a}+q_{b}U(1)_{b}+q_{c}U(1)_{c}
\ee
As regards the two extra anomalous $U(1)$'s, they can be cured by 
the generalized Green-Schwarz mechanism. 
Intriguingly, anomalous $U(1)$ are impossible to be consistently included in gauge theories 
while they
can be opportunely cured in string theories. 
 In the stringy extension of the (MS)SM as the ones in consideration, two new vector bosons $Z',Z''$ 
 are generically predicted, getting a mass through 
 a St\"uckelberg mechanism, and interacting through generalized Chern-
Simon (GCS) terms. In fact GCS are introduced in order to cancel all anomalies.
See \cite{Stuck1,Stuck2,Stuck3,Stuck5,Stuck9,Stuck10,Stuck11,Bianchi:2007fx} for an extensive 
discussion on these aspects.

In a minimal tree-nodes configuration, 
(MS)SM superfields can be reconstructed as 
follows:
$Q_{L}$  as $(a,\bar{b})$ or $(a,b)$; $U_{R}$ as $(\bar{a},\bar{c})$;
$D_{R}$ as $(\bar{a},c)$;
$H_{u}$ as $(b,c)$;
$H_{d}$ as $(\bar{b},c)$;
$L$ as $(\bar{b},\bar{c})$ or $(\bar{b},\bar{c})$;
$E_{R}$ as $(c,c')$.

For these intersections, the correspondent hypercharge is 
$U(1)_{Y}=\frac{1}{6}U(1)_{a}+\frac{1}{2}U(1)_{c}$. 

However, in order to guarantee that such a model
is free by tadpoles and that $U(1)_{Y}$ associated to $Y$ is 
a massless combination, usually one has to add extra exotic matter
in order to satisfy these two stringent condition. 
A complete classification of extra massive states was shown in \cite{DB18}.
Typically, extra vector-like pairs and charged singlets are
often introduced for consistency. 

These models cannot reproduce all the MSSM Yukawa couplings at perturbative level.
For example, $y_{u}H_{u}QU^{c}$ is generate at perturbative level,
but not $y_{d}H_{d}QD^{c}$ and $y_{l}H_{d}LE^{c}$.
However, they can be generated by $E2$-instantons,
opportunely intersecting D6-branes' stacks,
with $y_{d}\sim e^{-S_{E_{2}^{d}}}$ and $y_{l}\sim e^{-S_{E_{2}^{l}}}$ \cite{Blumenhagen:2006xt,Ibanez1,Ibanez2},
where $S_{E_{2}^{d},E_{2}^{i}}$
depend on geometric moduli associated to 3-cycles wrapped on $CY_{3}$ by $E_{2}^{d},E_{2}^{l}$-instantons.
Also a $\mu$-term is not generated at perturbative level, but it can be generated by exotic instantons 
as just proposed in \cite{Blumenhagen:2006xt,Ibanez1,Ibanez2}. In particular, as shown in  \cite{Blumenhagen:2006xt,Ibanez1}, one obtain 
 $\mu=e^{-S_{E2''}}M_{S}$
where $M_{S}$ is the string scale, while $e^{-S_{E2''}}$ depends by geometric moduli 
parameterizing 3-cycles wrapped on $CY_{3}$ by $E2''$-branes respectively. 
Practically, in these local models, we can consider $e^{-S_{E2'',E2^{d,l}}}$ as free-parameters,
depending on the particular geometry of the exotic instantons considered. 

Let us consider in this class of models, the presence of a 
new $E2$-brane
intersecting two times the $U(3)$-stack, 
two times the $U(1)$-stack, four time the $U(1)'$-stack. 
These mixed disk amplitudes lead to effective interactions between 
$U^{c},D^{c}$ and fermionic zero moduli (modulini).
In fact, modulini are obtained by open strings 
attached to $D6$-stacks and $E2$-stacks rather
than to $D6$-$D6$. Let us assume that this $E2$-instanton has 
a Chan-Paton factor $O(1)$, {\it i.e} it is placed on a $\Omega^{+}$-plane (symmetric). 
Calling $\tau^{i}$ modulini living between $U(3)$-$E2$, 
$\alpha$ modulini between $U(1)-E2$ and $\beta$ between $U(1)'-E2$,
the following effective interactions are generated: 
\be{effectivemod}
\mathcal{L}_{eff}\sim c_{f}^{(1)}U_{f}^{i}\tau_{i}\alpha+c_{f}^{(2)}D_{f}^{i}\tau_{i}\beta 
\ee
Integrating on the modulini space associated to the D6-E2 intersections, 
we obtain 
\be{obtain}
\mathcal{W}_{E2}=\int d^{6}\tau d^{4}\beta d^{2}\alpha e^{\mathcal{L}_{eff}}\ee
$$= \mathcal{Y}^{(1)}\frac{e^{-S_{E2}}}{M_{S}^{3}}\epsilon_{ijk}\epsilon_{i'j'k'}U^{i}D^{j}D^{k}U^{i'}D^{j'}D^{k'}$$
where $\mathcal{Y}^{(1)}_{f_{1}f_{2}f_{3}f_{4}f_{5}f_{5}}=c^{(1)}_{f_{1}}c^{(1)}_{f_{2}}c^{(2)}_{f_{3}}c^{(2)}_{f_{4}}c^{(2)}_{f_{5}}c^{(2)}_{f_{6}}$ is the flavor matrix, 
and $c^{(1,2)}$ are related on the particular homology and topology of the mixed disk amplitudes,
so that we can assume these as free parameters. 
Superpotential (\ref{obtain}) corresponds to the ope for $n-\bar{n}$ transitions
\be{ope}
\mathcal{O}_{n\bar{n}}=\frac{y_{1}}{\mathcal{M}_{E2}^{3}M_{SUSY}^{2}}(u^{c}d^{c}d^{d})(u^{c}d^{c}d^{c})
\ee
where $\mathcal{M}_{E2}^{3}=e^{+S_{E2}}M_{S}^{3}$, and $M_{SUSY}$ comes from the 
susy conversion of squarks into quarks through the exchange of a gaugino as a gluino, zino or photino,
$y_{1}=\mathcal{Y}^{(1)}_{111111}$.

For the quiver shown in Fig.1, a first problem is the perturbative generation 
of  $\mathcal{W}_{\Delta L=1}=y_{LQU}LQU^{c}$, violating $L$ of $\Delta L=1$. 
This superpotential has to be tuned close to zero, in order to avoid 
a dangerous proton destabilization. 

Now, let us discuss another possible case with four nodes, known 
as Madrid-embedding, with hypercharge 
$U(1)_{Y}=\frac{1}{6}U(1)_{a}+\frac{1}{2}U(1)_{c}+\frac{1}{2}U(1)_{d}$.
In this class of quiver, one could obtain discrete symmetries like 
R-parity from the Stueckelberg mechanism of anomalous $U(1)$s.
as discussed in \cite{Anastasopoulos:2015bxa}. 
For this motivation, the generation of a Majorana mass for the neutron,
as well as a Majorana mass for the neutrino, comes
from an opportune $E2$-instantons, 
dynamically breaking R-parity 
so that a selection rule $\Delta B=\Delta L=2$ will emerge. 
As a consequence, no other bilinear or trilinear R-parity violating superpotentials
are generated. 

A Madrid-embedding allows for 
$Q_{L}$ as $(a,\bar{b})$ or $(a,b)$;
$U_{R}$ as $(\bar{a},\bar{c})$ or $(\bar{a},\bar{d})$;
$D_{R}$ as $(\bar{a},c)$ or $(\bar{a},d)$;
$L_{L}$ as $(b,\bar{c})$, $(\bar{b},\bar{c})$, $(b,\bar{d})$, $(\bar{b},\bar{d})$;
$E_{R}$ as $(c,d)$ or $A_{c}$ or $S_{c}$;
$N_{R}$ as $A_{b}$ or $\bar{A}_{b}$ or $(c,\bar{d})$ or $(\bar{b},d)$;
$H_{u}$ as $(\bar{b},c)$ or $(b,c)$ or $(\bar{b},d)$ or $(b,d)$;
$H_{d}$ as $(b,\bar{c})$ or $(\bar{b},\bar{c})$ or $(b,\bar{d})$ or $(\bar{b},\bar{d})$.

Generically, also in this case, several MSSM Yukawa couplings will be not generated at pertubative level,
but they can be non-perturbatively generated by opportune $E2$-instantons, 
as mentioned above for three-nodes' quivers.

In Fig.2, we show a possible example, in which MSSM Yukawa couplings 
have to be generated by exotic instantons. 
In order to cancel all tadpoles and to guarantee a massless hypercharge, 
one has to introduce extra exotic matter, as in the case mentioned above.
A complete classification of exotic superfields introduced for consistency was shown in \cite{DB18}
also for this case. 
In Fig.2, we show how an $E2$-instanton generating (\ref{obtain})
 can be easily introduced as in the three node case. 

These examples seem to sustain the quite generality of such a mechanism,
for several models in literature \cite{DB1,DB23}.
 
In other models a possible viable alternative can be considered:
an $E2$-instanton,
with modulini
coupled to $Q$ and $D$ rather than $U$ and $D$,
as
 (\ref{obtain})
\be{effectivemod2}
\mathcal{L}_{eff}\sim c_{f}^{(3)}Q_{f,\alpha}^{i}\gamma_{i}\eta^{\alpha}+c_{f}^{(4)}D_{f}^{i}\gamma_{i}\zeta
\ee
where $\gamma_{i}$ are modulini living between $E2-U(3)$, $\eta$ between $E2-U(2)$, $\zeta$ between $E2-U^{I}(1)$
($U(1)^{I}$ is the involved $U(1)$ depending by the specific quiver). 
Integrating on the modulini space we generate an effective superpotential 
\be{obtain2}
\mathcal{W}_{E2}=\int d^{6}\tau d^{2}\zeta d^{8}\eta e^{\mathcal{L}_{eff}}
\ee
$$= \mathcal{Y}^{(2)}\frac{e^{-S_{E2}}}{M_{S}^{3}}\epsilon_{ijk}\epsilon_{i'j'k'}\epsilon^{\alpha\alpha'}\epsilon^{\beta\beta'}Q^{i}_{\alpha}Q_{\beta}^{j}D^{k}Q^{i'}_{\alpha'}Q^{j'}_{\beta'}D^{k'}$$
The associates ope for $n-\bar{n}$ transitions is
\be{ope}
\mathcal{O}_{n\bar{n}}=\frac{y_{2}}{\mathcal{M}_{E2}^{3}M_{SUSY}^{2}}(q^{c}q^{c}d^{d})(q^{c}q^{c}d^{c})
\ee
where again $\mathcal{M}_{E2}^{3}=e^{+S_{E2}}M_{S}^{3}$ and $y_{2}=\mathcal{Y}^{(2)}_{111111}$.

Finally, similarly to Fig.1-2, the mechanism can be implemented in a completely consistent quiver 
without extra exotic colored or electroweak states (see Fig.3) and with all SM yukawa couplings 
pertubatively allowed, 

The hypercharge in this model is the combination of $3$ charges, 
coming from $U(1)_{3}$, $U(1)$ and $U'(1)$:
\begin{equation}\label{Yper}
Y(Q)=c_3q_{3}+c_1q_{1}+c'_1q'_{1}
\end{equation}
We can fix the coefficients in such a way as to 
recover the standard hypercharges:
\begin{equation}\label{con1}
Y(Q)=\frac{1}{3}=c_3
\end{equation}
\begin{equation}\label{cond2}
Y(U^{c})=-\frac{4}{3}=-c_3-c'_1
\end{equation}
\begin{equation}\label{cond3}
Y(D^{c})=\frac{2}{3}=-c_3+c'_1
\end{equation}
\begin{equation}\label{cond4}
Y(L)=-1=c_1
\end{equation}
\begin{equation}\label{cond5}
Y(H_{d})=-Y(H_{u})=-1=-c'_1
\end{equation}
\begin{equation}\label{cond6}
Y(E^{c})=2=-c_1+c'_1
\end{equation}
\begin{equation}\label{cond6}
Y(N_{R})=0=-c_1-c'_1
\end{equation}
leading to 
\begin{equation}\label{result1}
c_3=\frac{1}{3},\,\,\,c_1=-1,\,\,\,c'_1=1 \quad \rightarrow \quad  Y=\frac{1}{3}q_{3}-q_{1}+q_{1'}
\end{equation}

Let us note that the quiver in Fig.~3 is free of tadpoles and the hypercharge $U(1)_{Y}$ is massless. 
A generic quivers has to satisfy two conditions 
in order to be anomalies'/tadpoles' free and 
in order to have a massless hypercharge.
The first one associated to tadpoles' cancellations is 
\be{condition1}
\sum_{a}N_a(\pi_{a}+\pi_{a'})=4\pi_{\Omega}
\ee
where $a=3,1,1'$ in the present case, $\pi_{a}$ 3-cycles wrapped by ``ordinary" D6-branes and $\pi_{a'}$ 3-cycles wrapped by the ''image" D6-branes. 
Eq.(\ref{condition1}) can be rewritten  
as a function of field representations 
\be{condition1a}
\forall \: a\neq a' \qquad \#F_a-\#\bar{F}_a+(N_{a}+4)(\# S_{a}-\# \bar{S}_{a})+(N_{a}-4)(\# A_{a}-\# \bar{A}_{a})=0\ee
where $F_a,\bar{F}_a,S_{a},\bar{S}_{a}, A_{a}, \bar{A}_{a}$ are fundamental, symmetric, antisymmetric
of $U(N_{a})$ and their conjugate.
For $N_{a}>1$ these coincide with the absence of irreducible $SU(N_{a})^{3}$ triangle anomalies. 
For $N_a=1$, these are stringy conditions that can be rephrased as absence of `irreducible' $U(1)^3$ {\it i.e.~} those arising from inserting all the vector bosons of the same $U(1)$ on the same boundary. 
Let us explicitly check tadpole cancellation for the $3,1,1'$ nodes:
\be{nodethree}
U(3):\,\,\,\,\,2 n_{Q}-n_{D}-n_{U}=6-3-3=0
\ee
\be{nodeuno}
U(1):\,\,\,\,\,2 n_{L}-n_{E}-n_{N}=6-3-3=0
\ee
\be{nodeunop}
U(1)':\,\,\,\,\,n_{E}-n_{N}+3 n_{D}-3 n_{U}=3-3+3{\cdot}3-3{\cdot}3-3=0 
\ee

The quiver in Fig.3 also 
satisfies the condition a massless vector boson $U(1)_{Y}$, with $Y=\sum_{a}c_{a}Q_{a}$,
\be{condition2}
\sum_{a}c_{a}N_{a}(\pi_{a}-\pi'_{a})=0
\ee
corresponding to 
 \be{condition3a}
c_{a}N_{a}(\#S_{a}-\# \bar{S}_{a}+\# A_{a}-\# \bar{A}_{a})
 \ee
 $$-\sum_{b\neq a} c_{b}N_{b}(\#(F_a,\bar{F}_b) - \#(F_a,F_b))=0$$
For $c^Y_3=1/{3}$, $c^Y_1=-1$, $c^{'Y}_1=1$, these conditions are satisfied in each quiver nodes. 
On the other hand, the massive (anomalous) $U(1)$'s correspond to  $3Q_3 + Q_1$ and to $3Q_3 - Q'_1$. 
This can can be demonstrated calculating the anomaly polynomial.

\subsection{Comments on proton decay, dark matter, FCNCs, Baryogenesis}

In our models, exotic instantons generate
 $\Delta L=2$ violating mass terms for neutrini
 and $\Delta B=2$ violating mass term for the neutron. 
As a consequence, in R-parity preserving models, R-parity is dynamically violated at non-perturbative level, 
so that other $\Delta L,B=1$ superpotentials are not generated.
As a consequence, a selection rule $\Delta B,L=2$ has emerged in this mechanism. 
No-proton decays operators have been generated.
In fact, 
\be{eff1}
\mathcal{W}_{MSSM}+\frac{1}{2}m_{N}NN+\frac{1}{\mathcal{M}_{n\bar{n}}^{3}}U^{c}D^{c}D^{c}U^{c}D^{c}D^{c}
\ee
cannot generate a proton decay as well as the lightest neutralino remains stable. 
Clearly the same is valid for 
\be{eff2}
\mathcal{W}_{MSSM}+\frac{1}{2}m_{N}NN+\frac{1}{\mathcal{M}_{n\bar{n}}^{3}}QQD^{c}QQD^{c}
\ee
On the other hand, extra diagrams CP-violating FCNCs in mesons' physics, such as $K_{0}-\bar{K}_{0}, B_{0}-\bar{B}_{0},...$,  are strongly suppressed in our model, up to $1000\, \rm TeV$-scale.
This is an interesting difference with respect to other neutron-antineutron models, 
usually involving extra colored states leading to non-negligible contributions to FCNCs. 
 
The next answer regards baryogenesis.
In fact, $(u^{c}d^{c}d^{c})^{2}/\mathcal{M}_{n\bar{n}}^{5}$
or $(qqd^{c})^{2}/\mathcal{M}_{n\bar{n}}^{5}$
can generate collisions $ud^{c}d^{c}\rightarrow \bar{u}^{c}\bar{d}^{c}\bar{d}^{c}$
or $qqd^{c}\rightarrow \bar{q}\bar{q}\bar{d}^{c}$.
Supposing an initial $\Delta L$ generated by RH-neutrini decays 
one has to be careful about washing-out action of six-quarks' collisions $B-L$
and sphalerons $B+L$. However, the problem is solved
if simply $m_{N}<\mathcal{M}_{n\bar{n}}$:
in this way,  six quarks collisions are not relevant 
during out-of-equilibrium RH-neutrini decays. 

Taking into account these considerations, let us discuss the different branches of parameters 
in our model. 
In order to obtain a $\mathcal{M}_{n\bar{n}}\simeq 1000\, \rm TeV$, 
we have several region of parameters. We discuss the main interesting ones. 

1) $M_{SUSY}\simeq \mathcal{M}_{E2}\simeq 1000\, \rm TeV$. 
$\mathcal{M}_{E2}=e^{+S_{E2}/3}M_{S}$, so that a case 
$M_{SUSY}\simeq M_{S}\simeq 1000\, \rm TeV$, with $e^{+S_{E2}/3}\simeq 1$
\footnote{However, in this limit, 3-cycles are so small that semiclassical approximation cannot be applied. 
Limit on semiclassical validity is $e^{-1/g_{s}}$. On the other hand, $g_{s}$ can also be large as $g_{s}\simeq 0.5$ or so.
},
seems a natural possibility. The condition $e^{+S_{E2}/3}$ is geometrically understood
as small radii 3-cycles wrapped by the $E2$-brane in the $CY_{3}$. 
In this scenario, $M_{N}\simeq e^{-S_{E2'}}\times (1000\, \rm TeV)$. 
However, usually, for a successful baryogenesis, 
$M_{N}$ mass has to be higher than Davidson-Ibarra bound 
$M_{N}>10^{6}\, \rm TeV$. 
On the other hand, RH neutrini masses are generated by $E2'$
and their values depend on the particular geometry of the mixed disk amplitudes.
As a consequence, a {\it resonant leptogenesis} scenario seems favored by 
this space of the parameters. considering at least two highly degenerate RH neutrini masses. 
In this case, Davidson-Ibarra bound is completely avoided, and RH neutrini masses 
can be also $M_{R}\simeq 1\, \rm TeV$ or so, as quantitatively shown in \cite{P1}. 
A RH neutrini degenerate spectrum can be understood geometrically 
by mixed disk amplitudes involved. 
As a consequence, $e^{-S_{E2'}}$ is constrained to $1\div 10^{-3}$. 
Hierarchy problem of the Higgs mass is alleviated of a $10^{-28}$ 
factor in this scenario, as $(M_{S}/10^{19}\, \rm GeV)^{2}$. 
In this scenario, LHC will not observe any signatures. 

2) $M_{SUSY}\simeq 1\, \rm TeV$ and $\mathcal{M}_{E}\simeq 10^{5}\, \rm TeV$.
In this case, a scenario for supersymmetry at LHC can be immediately tested 
in the next run. Such a scenario corresponds to several different
String-scales: $\mathcal{M}_{E}=M_{S}e^{+S_{E_{2}}/3}\simeq 10^{5}\, \rm TeV$ 
is compatible with $M_{S}=10\, \rm TeV$ and $e^{+S_{E2}/3}\simeq 10^{4}$
(large 3-cycles),
as well as with $M_{S}=10^{5}\, \rm TeV$ and $e^{+S_{E2}/3}\simeq 1$ (small 3-cycles). 
As regard leptogenesis, $M_{N}=e^{-S_{E2'}}M_{S}$ so that $M_{N}=1\div 10^{3}\, \rm TeV$
for several different combination of $e^{-S_{E2'}}$ and $M_{S}$ ({\it i.e} $e^{+E2/3}$). 
Again, in all these scenari, a resonant leptogenesis with at least two degenerate species of RH neutrini
is desired. 
In this case a stable neutralino dark matter with mass $10\div 1000\, \rm GeV$ is easily obtained,
and the mu-problem is solved and understood as a hierarchy generated by the $E2''$-brane
among the string scale and $\mu$ as $\mu=e^{-S_{E2''}}M_{S}$. 
\footnote{Alternatively, the presence of a parallel intersecting D-branes' world 
can contribute to dark matter. If the vev in the parallel sector was different
from the vev in our sector, the dark halo would be a non-collisional one, composed 
of dark atoms. See \cite{Addazi:2015cua} for a recent discussion on these aspects 
and implications in dark matter direct detection phenomenology. }

The most optimistic scenario for LHC is for $M_{S}=10\, \rm TeV$. In this case, 
signatures of Stringy Regge states will be immediately tested by LHC. 
For $M_{S}=10\, \rm TeV$, also signatures by anomalous $Z',Z''$ 
with Stueckelberg masses $m_{Z',Z''}\sim M_{S}$ can be 
tested at LHC, with peculiar channels from Generalized-Chern-Simons terms.
See \cite{StringyLHC} for a recent discussion on these aspects.

\section{Conclusions and remarks}

In this paper, we have discussed a simple mechanism 
directly generating a Majorana mass for the neutron
from exotic instantons. These effects are completely 
calculable and controllable.
Usually, a Weinberg operator is UV completed 
by massive ordinary fields, integrated-out 
at low energy limit. In our case, 
we have shown a counter-example 
to the Wilsonian UV completion:
a six quark operator like $(udd)^{2}/M_{n\bar{n}}^{5}$
is UV completed by exotic instantons
\footnote{
We are tempted to suggest that 
exotic instantons can be viewed as 
classicalons in internal dimension from the point of view of scattering amplitudes.
Classicalization was firstly suggested by Dvali and collaborators \cite{Dvali1,Dvali4}, 
and recently it was studied considered in the contest of 
non-local QFTs \cite{Addazi:2015ppa} (see also \cite{Addazi:2015dxa} for discussions of scattering amplitudes in $\mathcal{N}=1$ non-local QFTs). However, this conjecture will deserve future investigations beyond the 
purposes of these paper.}.

We have explicitly discussed examples of quivers
reproducing the (MS)SM, with a Majorana neutron and 
Majorana neutrini. In these models, 
operators leading to a proton decay are not generated.
In this framework, dark matter is not destabilized, 
and a successful resonant leptogenesis through RH neutrini decays 
can be realized. Starting from a B/L preserving model, 
Exotic instantons have dynamically broken R-parity, 
so that $\Delta (B-L)=2$ selection rules have naturally emerged, rather than 
imposed {\it ad hoc}. 
In the 'LHC era', LHC data will provide important inputs
for our model, constraining or likely discovering 
direct signatures also connected to our class of models, 
for several different regions of the parameters' space...
On the other hand, the possibility to improve the best current limit 
on neutron-antineutron transition in the near future is 
technically possible and 
well motivated by theoretical principles. 
The 'crazy idea' of Majorana in '37' has never been 
so relevant and intriguing as now!

\vspace{1cm} 

{\large \bf Acknowledgments} 
\vspace{3mm}

I would like to thank Massimo Bianchi, Ralph Blumenaghen, Gia Dvali, Cesar Gomez, Ryan Griffiths, Fernando Quevedo, Jim Halverson, Augusto Sagnotti, Jos\'e Valle
and the anonymous referee for useful comments and remarks. 
I also would like to thank the organizers of 
String Phenomenology 2015 (Madrid)
for hospitality, where this letter was started and 
 Ludwig Maximillian Universit\"at (M\"unchen)
for the hospitality during the completion of this letter. 
My work was supported in part by the MIUR research grant Theoretical Astroparticle Physics PRIN 2012CP-PYP7 and by SdC Progetto speciale Multiasse La Societ\'a della Conoscenza in Abruzzo PO FSE Abruzzo 2007-2013.

\end{document}